\documentclass[twocolumn,amsmath,amssymb,aps,superscriptaddress]{revtex4-2}                    

\usepackage{titlesec}
\usepackage[colorlinks,bookmarks=false,citecolor=blue,linkcolor=red,urlcolor=blue]{hyperref}
\usepackage{epsfig}
\usepackage{color}
\usepackage{subfigure}
\usepackage{graphicx}    
\usepackage{dcolumn}     
\usepackage{lipsum}      
\usepackage{braket}      
\input{insbox}

\newcommand{\be}{\begin{equation}}
\newcommand{\ee}{\end{equation}}

\definecolor{drkgr}{rgb}{0.05,0.6,0.2}

\begin{document}

\title{Relevance of on-site and intersite Coulomb interactions
                   in the Kitaev-Heisenberg magnet Na$_3$Co$_2$SbO$_6$}

\author{Pritam Bhattacharyya}
\email [Contact author: ] {pritambhattacharyya01@gmail.com}
\affiliation{Institute for Theoretical Solid State Physics, Leibniz IFW Dresden, Helmholtzstra{\ss}e~20, 01069 Dresden, Germany}
\affiliation{Department of Physics, Karpagam Academy of Higher Education, Coimbatore 641021, Tamil Nadu, India}
\affiliation{Centre for Computational Physics, Karpagam Academy of Higher Education, Coimbatore 641021, Tamil Nadu, India}

\author{Abdul Basit}
\affiliation{School of Physics, University of Melbourne, Parkville, VIC 3010, Australia}

\author{Thorben Petersen}
\affiliation{Institute for Theoretical Solid State Physics, Leibniz IFW Dresden, Helmholtzstra{\ss}e~20, 01069 Dresden, Germany}

\author{Stephan Rachel}
\affiliation{School of Physics, University of Melbourne, Parkville, VIC 3010, Australia}

\author{Satoshi Nishimoto}
\affiliation{Institute for Theoretical Solid State Physics, Leibniz IFW Dresden, Helmholtzstra{\ss}e~20, 01069 Dresden, Germany}
\affiliation{Department of Physics, Technical University Dresden, 01069 Dresden, Germany}

\author{Liviu Hozoi}
\email [Contact author: ] {l.hozoi@ifw-dresden.de}
\affiliation{Institute for Theoretical Solid State Physics, Leibniz IFW Dresden, Helmholtzstra{\ss}e~20, 01069 Dresden, Germany}

\begin{abstract}
\noindent
The detection of considerable spin frustration in honeycomb cobalt oxide compounds indicates the
presence of sizable Kitaev interactions in these systems, enlarging the pool of Kitaev spin liquid
candidates.
Several key questions remain to be answered, as basic as the mechanisms behind Kitaev couplings in
Co$^{2+}$ $t_{2g}^5e_g^2$ magnets.
Analyzing the quantum chemistry of interacting magnetic moments in Na$_3$Co$_2$SbO$_6$, a representative
$LS$-coupled $t_{2g}^5e_g^2$ oxide, we find that the Kitaev and off-diagonal $\Gamma$ interactions
are substantial and antiferromagnetic but somewhat weaker than the Heisenberg contribution.
%
All nearest-neighbor couplings feature massive contributions from direct Coulomb exchange and/or on-site
multiconfigurational dressing, mechanisms not considered so far in descriptive models of Kitaev-Heisenberg
magnetism.
%
These findings call for systematic wave-function quantum chemical studies in order to understand
direct-indirect exchange synergies in Kitaev-Heisenberg magnets and how to possibly tune intersite
couplings towards the Kitaev spin liquid ground state.
\end{abstract}

\date\today
\maketitle
{\it Introduction.}
Science relies on models.
First hypothesized, then tested, a model may latter on turn valid or inadequate.
Atomic models, for example, have gone through many changes over time, culminating with quantum theory
of atoms.
Famed byproducts of the latter are the notions of spin and exchange.

%
Exchange is ubiquitous in electronic matter.
It can be direct, as inferred in the 1920s, but also indirect, i.e., proceeding through inter-atomic
charge transfer (CT).
The functions of magnetic materials in technological applications rely all on exchange.

Exchange can be isotropic as in Heisenberg magnets or, when spin-orbit couplings (SOCs) are present,
highly anisotropic as in Ising or more recently identified Kitaev magnetic systems.
Proposed initially as an exactly soluble magnetic model, Kitaev's construct \cite{Kitaev2006} became
a major reference point in condensed matter magnetism.
It is of interest not only as fundamental research but also for possible applications in quantum
technologies.

Direct, Coulomb exchange is computed exactly in {\it ab initio} wave-function-based quantum chemistry
\cite{olsen_bible,fulde_book} but not in density functional theory (DFT) simulations relying on
functionals typically employed in solid state physics, local density and generalized gradient
functionals.
In current descriptive magnetic interaction models, direct Coulomb exchange is ignored --- not only
for the case of edge-sharing Kitaev-Heisenberg magnets \cite{Khaliullin_2005,Jackeli_2009,kitaev_review_2017,
Liu_2018,Liu_2020,Sano_2018,Winter_2022,Liu_2023} but also in e.\,g. edge-sharing cuprate chains, another
class of emblematic frustrated magnets.
Direct Coulomb exchange is of particular relevance to edge-sharing connectivity since certain pairs of
metal (M) $d$ orbitals are in `direct contact' through the space left `empty' between the two bridging
ligands (Ls), different from corner-sharing ML$_6$ octahedra and linear M-L-M bonds.

As to excitations, i.\,e., L$\rightarrow$M CT (i.e., superexchange) and M$\rightarrow$M CT (hopping-mediated
kinetic exchange): those do not explicitly enter canonical DFT.
In non-DFT context, analytical work based on simplified valence-band effective CT models and 2nd-order
perturbation theory may\,
(i) imply remarkable intuition,
(ii) reveal fascinating new physics, and
(iii) open up exciting new research avenues;
this was indeed the case with the initial $LS$-coupled, CT valence-band model (and analysis) of Khaliullin
and Jackeli \cite{Khaliullin_2005,Jackeli_2009}.
Yet, there is room (or even the need) for investigations beyond effective CT valence-band models
and simple, perturbation-theory expressions.
%
Further, if we want to understand the interplay of indirect CT exchange and direct Coulomb exchange,
we need to put both into the same machinery; as pointed out above, only the quantum chemical computational
machinery is presently able to describe both CT excitations and direct exchange.


Here we pin down the underlying exchange mechanisms in Na$_3$Co$_2$SbO$_6$, a honeycomb cobaltate
whose macroscopic magnetic properties indicate substantial frustration \cite{Songvilay_2020,Kim_2022,
Sanders_et_al,Li_et_al}, presumably arising from sizable, bond-dependent \cite{Jackeli_2009,
Liu_2018} anisotropic intersite interactions.
We first demonstrate the capabilities of our quantum chemical methodology through a scan
of the many-body Co-site multiplet structure, benchmarked against existing inelastic neutron scattering
(INS) measurements \cite{Kim_2022} and analysis of X-ray spectra \cite{Veenendaal_et_al}.
Focusing then on intersite effective couplings, we unveil the morphology of Co-Co ansiotropic exchange:
%
important nearest-neighbor interaction mechanisms turn out to be direct Coulomb exchange and dressing
with on-site excitations, according to the quantum chemical data.
Such physics being neglected in current descriptive electronic models for Kitaev-Heisenberg magnetism,
our work redefines the overall map of symmetric anisotropic pseudospin interactions in quantum matter.

  \

\begin{figure}[b]
\centering
\includegraphics[width=1.0\linewidth]{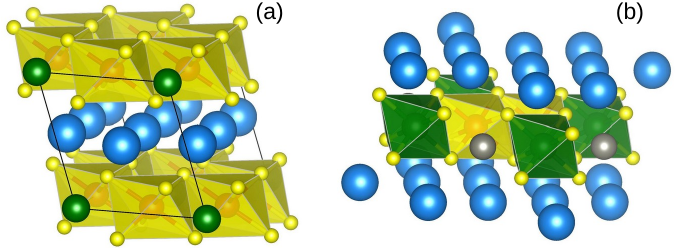}
\caption{
(a) Successive atomic layers in Na$_3$Co$_2$SbO$_6$.
The CoO$_6$ octahedra are represented in yellow; Na and Sb species are shown in blue and green, respectively.
Each Sb sits in the center of a hexagonal ring formed by Co ions.
(b) Cluster employed for deriving the nearest-neighbor effective magnetic couplings.
Also the SbO$_6$ octahedra are highlighted, in green.
The embedding is not pictured.
A different perspective is provided in Supplemental Material.
}
\label{fig_crystal_str}
\end{figure}


{\it Co-site multiplet structure.\,}
The distribution of the different atomic species in Na$_3$Co$_2$SbO$_6$ is illustrated in Fig.\;\ref{fig_crystal_str}.
The Co$^{2+}$ magnetic ions form a honeycomb network.
The valence electronic structure of the Co$^{2+}$ magnetic centers, as obtained by {\it ab initio}
quantum chemical analysis, is detailed in Table\;\ref{d5}.
Here we built on insights gained from quantum chemical investigations of a series of other cobaltates,
$d^6$ \cite{Hozoi_2009_d6}, $d^7$ \cite{Iakovleva_d7}, and $d^8$ \cite{Albert_d8}.
Various features concerning the Co-ion ground state and multiplet structure can be directly compared
with info extracted from spectroscopic investigations already carried out on Na$_3$Co$_2$SbO$_6$: the
degree of $t_{2g}^5e_g^2$--$t_{2g}^4e_g^3$ configurational mixing in the ground-state wave-function
\cite{Veenendaal_et_al}, the trigonal splitting of the Co 3$d$ $t_{2g}$ levels $\delta$ \cite{
Veenendaal_et_al}, and the position of the low-lying `$LS\delta$' exciton \cite{Kim_2022}.

\begin{table*}[th!]
\caption{
Co$^{2+}$ $3d^7$ multiplet structure in Na$_3$Co$_2$SbO$_6$.
SC stands for a single-configuration ($t_{2g}^5e_g^2$) $S\!=\!3/2$ calculation.
Each value in the last two columns indicates a Kramers doublet (KD);
for the $^4T$ terms, the KDs are listed as groups of closely spaced states.
Only states with relative energies lower than 2 eV are listed.
Notations as in $O_h$ symmetry are used \cite{Sugano}, though the actual point-group symmetry is
lower \cite{Viciu_2007}.
}
\begin{tabular}{lllllr}
\hline
\hline\\[-0.20cm]
Relative                       &SC      &CASSCF\footnote{Orbitals optimized for the lowest three $S\!=\!3/2$ roots; the SC splittings in the adjacent column are obtained using this orbital basis.}
						  &CASSCF\footnote{Orbitals optimized for all $S\!=\!3/2$ and the lowest 20 (out of 50) $S\!=\!1/2$ roots;
						                   all $S\!=\!3/2$ and the lowest 20 $S\!=\!1/2$ 3$d^7$ states were included in the spin-orbit treatment,
								   at both CASSCF and MRCI levels.}
                                                                     &MRCI\\
energies (eV)                  &        &         &+SOC              &+SOC\\
\hline
\\[-0.20cm]
$^4T_{1g}$ ($t_{2g}^5e_g^2$)   &0       &0\footnote{8\% $t_{2g}^4e_g^3$ character, as also estimated by van Veenendaal {\it et al.}~\cite{Veenendaal_et_al} from the analysis of X-ray spectra.}
                                                  &0\footnote{0.33\% admixture of excited state configurations through 2nd-order SOCs.}
                                                                     &0\\
			       &0.10    &0.06     &0.03, 0.07        \footnote{$j\!=\!3/2$ quartet in cubic symmetry.}
                                                                     &0.03\footnote{27.5 meV, in agreement with the experimentally observed exciton at 28--29 meV \cite{Kim_2022}.},
                                                                      0.07\\
			       &0.11    &0.06     &0.13, 0.14, 0.15  \footnote{$j\!=\!5/2$ sextet in cubic symmetry.}
                                                                     &0.13, 0.13, 0.15 \\[0.10cm]
$^4T_{2g}$ ($t_{2g}^4e_g^3$)\footnote{The $^4T_{1g}$ ($t_{2g}^4e_g^3$) levels lie at 2.9--3.05 eV.}
                               &        &0.85     &0.81, 0.82        &0.88, 0.89     \\
			       &        &0.87     &0.84, 0.85, 0.86
								     &0.91, 0.91, 0.92\\
                               &        &0.88     &0.88              &0.94           \\[0.10cm]
$^4\!A_{2g}$ ($t_{2g}^3e_g^4$) &        &1.83     &1.72, 1.72        &1.77, 1.83     \\[0.10cm]
$^2\!E_g$ ($t_{2g}^6e_g^1$)    &        &         &1.93, 1.98        &1.85, 1.85     \\[0.10cm]
\hline
\hline
\end{tabular}
\label{d5}
\end{table*}

To disentangle crystal-field effects, on-site Coulomb interactions, and SOCs, embedded-cluster
quantum chemical calculations were first performed at the single-configuration (SC) $t_{2g}^5e_g^2$
level, i.\,e., excluding other orbital occupations (see Supplemental Material, SM, for computational
details)
\footnote{See Supplemental Material at [URL], which includes Refs.
\cite{Viciu_2007,Molpro,Klintenberg_et_al,Derenzo_et_al,olsen_bible,MCSCF_Molpro,MRCI_Molpro,SOC_Molpro,Co_basis,O-ligands,Schautz1998,Sb_pseudo,Pierloot1995,Fuentealba_Na,PM_method,Yadav2016,Ir214_niko_15,Liu_2020}
for detailed information about the numerical calculations.}.
SC results are provided in the first column of Table\;\ref{d5}: it is seen that the $^4T_{1g}$
($t_{2g}^5e_g^2$) manifold is split by trigonal and residual lower-symmetry \cite{Viciu_2007} fields
into distinct components.
%
By allowing subsequently for all possible orbital occupations within the Co 3$d$ shell, which is
referred to as complete-active-space self-consistent-field (CASSCF) \cite{olsen_bible,MCSCF_Molpro},
an admixture of 8\% $t_{2g}^4e_g^3$ character is found in the ground-state CASSCF wave-function,
in agreement with conclusions drawn from the analysis of X-ray spectra \cite{Veenendaal_et_al}.
Interestingly, given the low point-group symmetry \cite{Viciu_2007}, the $t_{2g}^5e_g^2$--$t_{2g}^4e_g^3$
interaction implies also Coulomb matrix elements that in cubic environment are 0 by symmetry: the
trigonal splitting within the $^4T_{1g}$ manifold is consequently reduced from a bare value of 100
meV (SC results in Table\;\ref{d5}) to 60 meV (CASSCF data) --- while $^4T_{1g}$($t_{2g}^5e_g^2$)--$^4T_{1g}$($t_{2g}^4e_g^3$) interaction occurs already
in cubic environment, differential effects may appear within the group of `initial' $T_{1g}$ levels
in symmetry lower than $O_h$.
Such physics was not discussed so far in effective-model theory \cite{Kim_2022,Veenendaal_et_al,
Liu_2020,Winter_2022,Liu_2023}.
Significantly heavier `dressing' may occur in the case of multi-M-site, molecular-like $j\!\approx\!1/2$
\cite{Petersen_sci_rep} and $j\!\approx\!3/2$ \cite{Petersen_nat_commun} spin-orbit states, up to the
point where the picture of `dressing' even breaks down \cite{Petersen_nat_commun}.

Upon including SOCs, at either CASSCF or multireference configuration-interaction (MRCI) \cite{
olsen_bible,MRCI_Molpro} level, additional splittings occur.
The lowest on-site excitation is computed at 27.5 meV (see footnote f in Table\;\ref{d5}), in excellent
agreement with the outcome of INS measurements \cite{Kim_2022}.
It is seen that, for the lower part of the spectrum, the MRCI corrections to the CASSCF relative
energies are moderate.

{\it Magnetic interactions.}
For a block of two adjacent edge-sharing ML$_6$ octahedra in layered honeycomb magnets, the highest
possible point-group symmetry is $C_{2h}$.
This implies a generalized bilinear effective spin Hamiltonian of the following form for a pair of
adjacent 1/2-pseudospins ${\bf{\tilde{S}}}_i$ and ${\bf{\tilde{S}}}_j$\,:
\begin{equation}
\label{eqn:Hamil}
\mathcal{H}_{ij}^{(\gamma)} = J{\bf{\tilde{S}}}_i\cdot {\bf{\tilde{S}}}_j + \\
                              K\tilde{S}_i^{\gamma}\tilde{S}_j^{\gamma} + \\
                              \sum_{\alpha\neq\beta} \Gamma_{\alpha\beta} \\
                              (\tilde{S}_i^{\alpha}\tilde{S}_j^{\beta} +  \\
                              \tilde{S}_i^{\beta}\tilde{S}_j^{\alpha}).
\end{equation}
The $\Gamma_{\alpha\beta}$ coefficients denote the off-diagonal components of the 3$\times$3
symmetric-anisotropy exchange tensor, with $\alpha,\beta,\gamma\!\in\!\{x,y,z\}$.
For e.\,g.~a $z$-type M-M bond (i.\,e., M$_2$L$_2$ plaquette normal to the $z$ axis),
$\Gamma\!\equiv\!\Gamma_{xy}$ and $\Gamma'\!\equiv\!\Gamma_{yz}\!=\!\Gamma_{zx}$.

A number of studies of the INS spectra of Na$_3$Co$_2$SbO$_6$ arrive to
%
antiferromagnetic Kitaev coupling $K$ \cite{Kim_2022} plus sizable, antiferromagnetic off-diagonal $\Gamma$
\cite{Kim_2022,Li_et_al}, and indicate that an antiferromagnetic $K$ requires ferromagnetic Heisenberg
interaction with $K\!\sim\!|J|$ \cite{Sanders_et_al},
although fits with ferromagnetic $K$ are also available \cite{Kim_2022,Sanders_et_al}.
In fact, a ferromagnetic-$K$ model can lead to the identical magnon spectrum as an antiferromagnetic-$K$
model \cite{Sanders_et_al,Chaloupka_prb}.
That is, fitting an effective model to the experiment has limited evidential value.
A relatively large ferromagnetic Heisenberg $J$ is also proposed by analysis of effective models relying on
Co-Co kinetic exchange, Co-O$_2$-Co superexchange, and intersite hoppings extracted from DFT
computations \cite{Winter_2022}.

For an {\it ab initio} quantum chemical perspective, we scanned the nearest-neighbor interaction
landscape at the SC (i.e., $t_{2g}^5e_g^2$--$t_{2g}^5e_g^2$ Co nearest neighbors, no excited-state
configurations considered), CASSCF, and MRCI levels (see SM for technicalities).
This allows to distinguish between
(i) direct, Coulomb exchange (the only available channel at SC level),
(ii) Co-Co kinetic exchange (additionally accounted for in the CASSCF computation with all 3$d$
orbitals of the two Co sites considered in the active space), and
(iii) Co-O$_2$-Co superexchange (physics considered by MRCI).
Remarkably, for $J$, $K$, and $\Gamma$, we find that CASSCF (i.\,e., CT kinetic exchange)
and MRCI (CT superexchange)
bring only minor corrections  to the SC values (see Table\;\ref{couplings}).

\begin{table}[b]
\caption{
Magnetic couplings (meV) for the $C_{2h}$ Co-Co link \cite{Viciu_2007}.
The singlet, triplet, quintet, and septet associated with each of the possible ($3\!\times\!3$)
$t_{2g}^5e_g^2$--$t_{2g}^5e_g^2$ orbital occupations were included in the spin-orbit treatment, which
yields 144 spin-orbit states;
the lowest four were mapped onto the model of two interacting 1/2 pseudospins (\ref{eqn:Hamil}), as
described in \cite{Ir214_niko_15,Yadav2016}.
SSCAS stands for single-site complete-active-space (Co-Co CT excluded);
all possible $d$-$d$ excitations, on-site\;+\;intersite, are considered in CASSCF.
}
\begin{tabular}{l  c  c  c  r }
\hline
\hline\\[-0.20cm]

Method \hspace{0.35cm}
                  &\hspace{0.6cm} $J$ \hspace{0.6cm}
                             &\hspace{0.6cm} $K$ \hspace{0.6cm}
                                     &\hspace{0.6cm} $\Gamma$ \hspace{0.6cm}
                                                  &\hspace{0.6cm} $\Gamma^\prime$ \hspace{0.10cm}   \\

\hline\\[-0.15cm]

SC                &--1.41    &0.45   &0.55        &--0.16    \\
SSCAS             &--1.50    &0.57   &0.47        &0.24\\
CASSCF            &--1.30    &0.54   &0.50        &0.21\\
MRCI              &--1.18    &0.53   &0.51        &0.17     \\
\hline
\hline
\end{tabular}
\label{couplings}
\end{table}

\begin{figure}[b]
\centering
\includegraphics[width=1.0\linewidth]{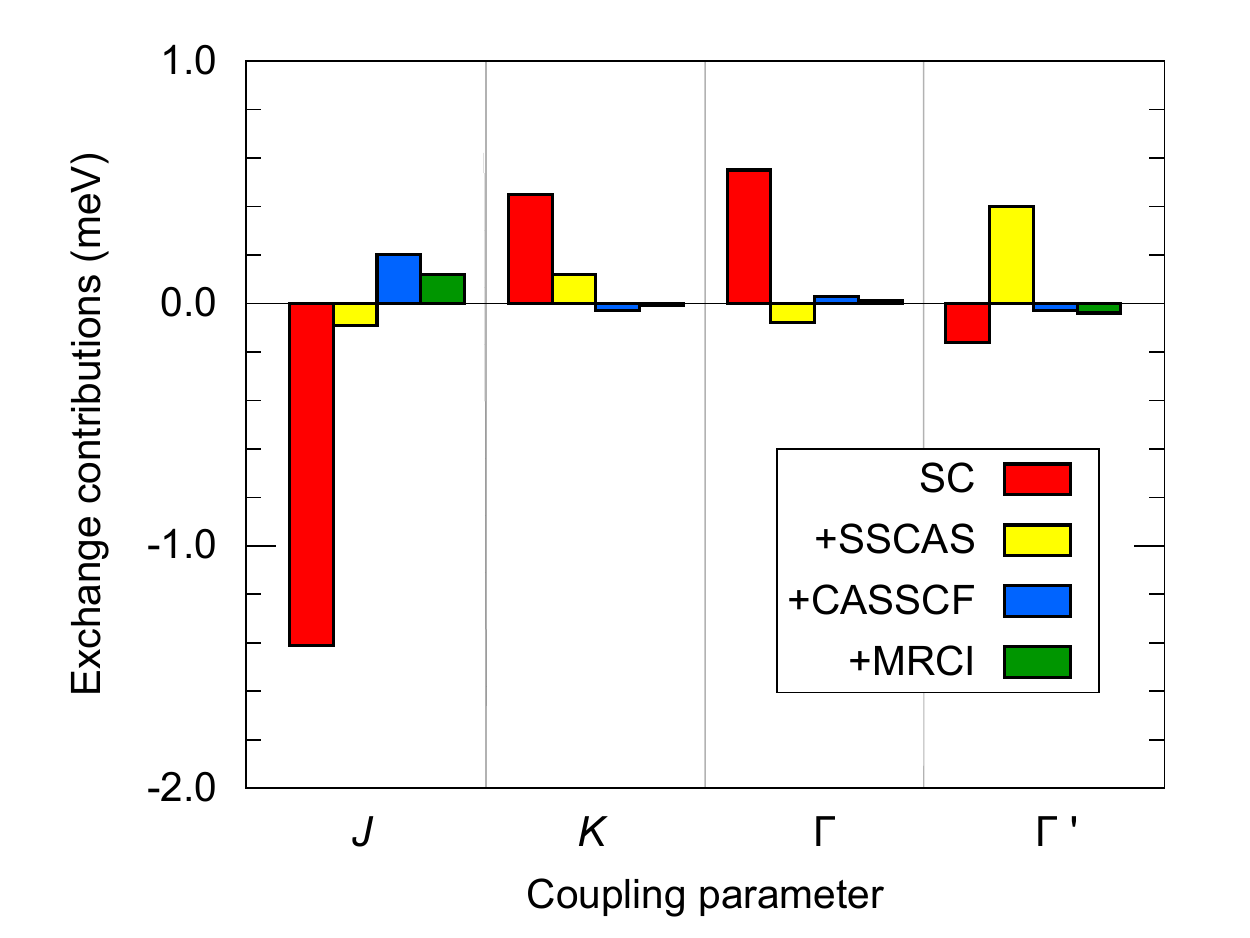}
\caption{
Direct exchange (red bars), CT kinetic exchange (blue), and CT superexchange plus additional
correlation effects (green) in Na$_3$Co$_2$SbO$_6$.
Yellow bars represent the effect of dressing with on-site excitations.
Red and yellow contributions are ignored in current descriptive exchange models.
}
\label{fig_NCSO_bars}
\end{figure}

Also interesting is the effect of on-site excitations, i.\,e., the admixture of $t_{2g}^4e_g^3$
character to the leading Co$^{2+}$ $t_{2g}^5e_g^2$ electron configuration (discussed as well in the
previous section), on $\Gamma'$.
As shown on the second line of Table\,\ref{couplings} and in Fig.\;\ref{fig_NCSO_bars}, this on-site
multiconfigurational `dressing' reverts the sign of $\Gamma'$, from ferromagnetic at SC level to
antiferromagnetic by single-site complete-active-space (SSCAS) multiconfiguration calculations where
Co-Co hopping is excluded.
The sign of $\Gamma'$ remains then positive (i.\,e., antiferromagnetic) when including additional
electronic excitations in the CASSCF and MRCI spin-orbit computations (the lowest two lines in
Table\;\ref{couplings}).
It turns out that, compared to the SC calculation, on-site Coulomb matrix elements
considered in the SSCAS numerical experiment modify the sequence of two of the four eigenstates in
the two-site magnetic problem.
This is how the sign of $\Gamma'$ changes.

\begin{table}[b]
\caption{
Nearest-neighbor effective magnetic couplings (meV) for the $C_{2h}$ Co-Co link \cite{Viciu_2007}
in $XXZ$ representation (see SM or e.\,g.~ref.~\cite{Winter_2022} for conversion relations).
}
\begin{tabular}{l  c  c  c  r }
\hline
\hline\\[-0.20cm]

Method \hspace{0.35cm}
                  &\hspace{0.45cm} $J_{xy}$ \hspace{0.45cm}
                             &\hspace{0.45cm} $J_z$ \hspace{0.45cm}
                                     &\hspace{0.6cm} $J_{\pm\pm}$ \hspace{0.45cm}
                                                  &\hspace{0.45cm} $J_{z\pm}$ \hspace{0.10cm}   \\

\hline\\[-0.15cm]

SC                &--1.34    &--1.11   &--0.31        &0.12   \\
SSCAS             &--1.63    &--0.68   &--0.17        &--0.16   \\
CASSCF            &--1.43    &--0.51   &--0.19        &--0.12   \\
MRCI              &--1.29    &--0.44   &--0.20        &--0.09   \\
\hline
\hline
\end{tabular}
\label{couplings_XXZ}
\end{table}

Transformed to $XXZ$ frame (see the discussion and conversion relations in SM), the nearest-neighbor
MRCI coupling parameters change to $J_{xy}\!=\!-1.29$, $J_z\!=\!-0.44$, $J_{\pm\pm}\!=-0.20$, and
$J_{z\pm}\!=\!-0.09$.
Their dependence on the various exchange mechanisms is illustrated in Table\;\ref{couplings_XXZ}:
it is seen that $J_{xy}$ is essentially determined by Coulomb exchange, while for the remaining nearest-neighbor
effective interactions also other contributions are significant, most of all, the dressing with on-site
excitations.
The relevance of the $XXZ$ effective spin model to the magnetism of Na$_3$Co$_2$SbO$_6$ is discussed
in ref.\;\cite{Li_PRB_2024}.

While the discussion has been focussed so far on the pair of edge-sharing CoO$_6$ octahedra displaying
$C_{2h}$ point-group symmetry \cite{Viciu_2007}, a similar fine structure is found for the excitation
spectrum of the lower-symmetry, $C_i$ Co$_2$O$_{10}$ unit:
the excitation energies of the lowest three excited states (defined by the interaction of the two 1/2
pseudospins) differ on average by 10\%.
Whether certain details in the experimental spectra can be explained by considering two different sets
of Co-Co magnetic links (i.\,e., two different sets of nearest-, second-, and third-neighbor couplings)
remains to be clarified in forthcoming work.
%


\begin{figure}[]
\centering
\includegraphics[width=0.85\columnwidth]{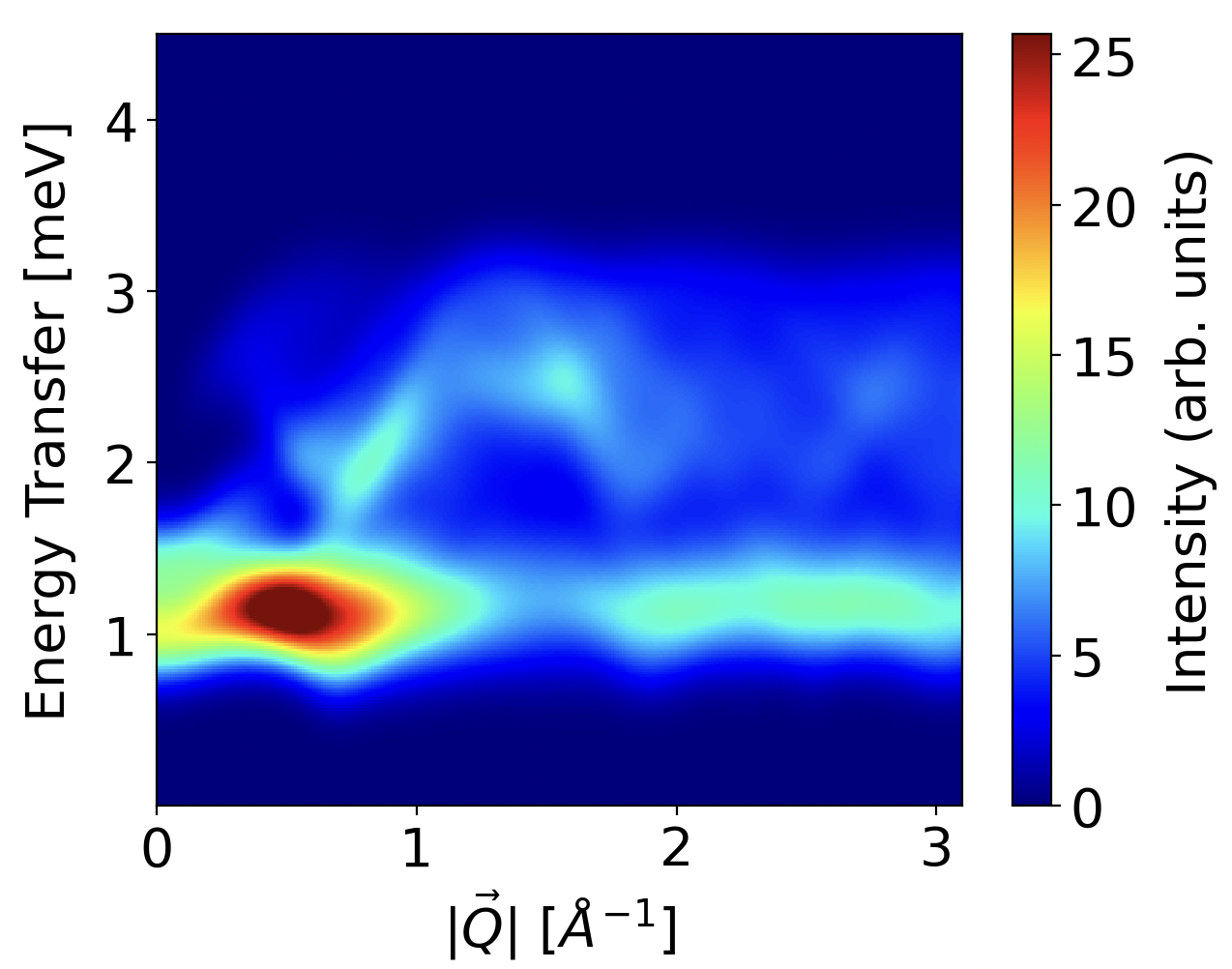}
\caption{
Powder-averaged dynamical structure factor for Na$_3$Co$_2$SbO$_6$.
Data were computed within linear spin-wave theory, using the nearest-neighbor MRCI parameters from
Table\,\ref{couplings} and $J_3\!=\!0.68$ meV
(details: Gaussian broadening of $\sigma=0.18$ and intensity cut-off at $5/6$ of maximum intensity).
}
\label{fig:structure-factor}
\end{figure}

The quantum chemical calculations were limited to adjacent CoO$_6$ magnetic units and
nearest-neighbor $K$, $J$, $\Gamma$, and $\Gamma'$ effective couplings.
%
%
%
Widely used to study magnetic ground states and excitations on Kitaev-Heisenberg honeycomb lattices
are effective spin models augmented with second- and third-neighbor $J_2$ and $J_3$ Heisenberg
interactions.
%
%
%
For Na$_3$Co$_2$SbO$_6$, the MRCI nearest-neighbor couplings alone yield ferromagnetic order, and
antiferromagnetic longer-ranged Heisenberg exchange ($J_2\!\simeq\!0.25$ or $J_3\!\simeq\!0.3$ or
combinations thereof) is required to tune to a zigzag ground state, in line with the literature and
also own exact-diagonalization calculations.
By employing linear spin-wave theory for the $K$-$J$-$\Gamma$-$\Gamma'$-$J_2$-$J_3$ model with
a classical zigzag ground state, we computed the dynamical structure factor
\begin{equation}\label{S_perp}
\mathcal{S}^{\perp}(\boldsymbol{Q},\omega) = \sum_{\alpha\beta} \left( \delta_{\alpha\beta} - \frac{Q^\alpha Q^\beta}{Q^2}\right) \mathcal{S}^{\alpha\beta}(\boldsymbol{Q},\omega)
\end{equation}
with the matrix elements
\begin{equation}\label{S_ab}
\mathcal{S}^{\alpha\beta} = \sum_{ij}\int \frac{d\tau}{2\pi} e^{-i\omega \tau} \left\langle S^\alpha_{-\boldsymbol{Q}_i}(0) S^\beta_{\boldsymbol{Q}_j}(\tau)\right\rangle\ .
\end{equation}
Here $S^\alpha_{\boldsymbol{Q}_i}$ is a Fourier-transformed spin operator in the Heisenberg picture,
and $i, j$ are lattice sites within the magnetic unit cell.
For Na$_3$Co$_2$SbO$_6$, the available INS data were measured on powder samples, and we thus considered
averaging over all momentum transfer directions.
The INS cross section then becomes $\propto\!F(Q)^2\int d\Omega \mathcal{S}^{\perp}(\boldsymbol{Q},\omega)$\,,
where $F(Q)$ is the form factor for Co$^{2+}$ \cite{form_factor}.
For the given MRCI effective couplings, the best match of the basic INS spectral features was found
for $J_2\!=\!0$ and $J_3\!=\!0.68$ meV, see Fig.\;\ref{fig:structure-factor};
small $J_2$ contributions would still lead to similar structure factors, but once $J_2$ reaches the
order of $J_3$ the main features start to differ.

With the exchange parameters used for the plot in Fig.\;\ref{fig:structure-factor}, the Curie-Weiss
temperature is calculated as $\Theta_{\rm CW}=-S(S+1)(J+2J_2+J_3+K/3)=2.8\;\text{K}$, indicating weak
ferromagnetic character.
This agrees with the experimentally determined $ab$-plane averaged $\Theta_{\rm CW} = 3.9\;\text{K}$
\cite{Yan_2019,Li_et_al}.
Several comments are in order.
While the successful synthesis of single crystals of Na$_3$Co$_2$SbO$_6$ was reported \cite{Yan_2019},
corresponding INS data are not available yet --- the published INS results were obtained on
polycrystalline samples.
What jumps out is that all three available experiments \cite{Kim_2022,Songvilay_2020,Sanders_et_al}
find strong intensity close to $Q\!=\!0.5$.
While strong signal is also found in our Fig.\,\ref{fig:structure-factor} around $Q\!=\!0.5$, the
corresponding energy is lower by roughly 1 meV and there is less intensity for the branch extending
towards $Q\!\approx\!1$.
This disagreement remains to be clarified by future work --- other effects that need to be addressed
are
%
%
cyclic exchange \cite{Janssen_PRL_2023,Janssen_PRB_2024,Li_PRB_2024,Co_multi_q_Gu_25,Li2024_et_al},
additional anisotropies due to the presence of two, symmetry-inequivalent sets of Co-Co links on a given
hexagonal ring \cite{Viciu_2007}, and second-neighbor antisymmetric interactions (allowed by symmetry).
%
That surprisingly small values for the ring-exchange interaction may have dramatic impact on the
magnetic excitation spectra has been recently shown for honeycomb CoTiO$_3$ \cite{Li2024_et_al}.
As concerns the additional anisotropies due to the presence of two different sets of Co-Co magnetic
bonds (i.\,e., intrinsic uniaxial deformation of the honeycomb lattice that removes $C_3$ rotation
symmetry at the Co sites) \cite{Viciu_2007}: this yields not only two different sets of nearest-,
second-, and third-neighbor couplings but also a $\Gamma''$ parameter for one set of Co-Co links and
a more complicated form of cyclic exchange.

{\it Conclusions.}\,
In sum, analyzing the quantum chemistry of
%
interacting $t_{2g}^5e_g^2$ magnetic moments, we identify intersite Coulomb exchange and on-site
multiconfigurational dressing (red and yellow bars in in Fig.\;\ref{fig_NCSO_bars}) as important
Kitaev-Heisenberg interaction channels.
The Coulomb exchange contributions to $K$, $J$, $\Gamma$, and $\Gamma'$ represent firm, assertive results:
obtaining those requires computations at the most basic level of approximation in {\it ab initio}
electronic-structure theory, Hartree-Fock-like, different from the much more sophisticated subsequent
calculations required to estimate the role of correlations/excitations.
Similar results on the magnitude of Coulomb exchange should be obtained by DFT computations with
functionals that incorporate exact (i.e., Hartree-Fock) exchange but disregard any correlations.
\footnote{
Describing kinetic exchange and superexchange (i.e., intersite {\it excitations}) through the (exchange-)correlation
functional remains however elusive.
}
As co-mechanism to intersite interactions, both isotropic and anisotropic, Coulomb exchange has already
been pointed out in quantum chemical studies of hexagonal $d^5$ RuCl$_3$ \cite{rucl3_naruo2} and
triangular-lattice $d^5$ NaRuO$_2$ \cite{rucl3_naruo2,Bhattacharyya_2023};
spotting it as
%
important interaction mechanism on a hexagonal $d^7$ lattice suggests that anisotropic Coulomb exchange
is ubiquitous in Kitaev-Heisenberg magnets.
Finally, the renormalization of intersite couplings through on-site multiconfigurational dressing is
an effect that might be also relevant to e.\,g.~$d^8$ NiX$_2$ nickelates \cite{Stavropoulos_prl}.
%
%
%

 \

{\it Data Availability.}
Research data related to this work have
been deposited in the RADAR database under the https://doi.org/10.22000/tny338gct87gzce4.

 \

{\it Acknowledgments.}
We thank A.~Tsirlin, R.~C.~Morrow, S.~L.~Drechsler, and M.~Richter for discussions and U.~Nitzsche
for technical support.
P.\,B., T.\,P., and L.\,H. acknowledge financial support from the German Research Foundation (Deutsche
Forschungsgemeinschaft, DFG), project number 468093414.
S.\,N.~acknowledges financial support through the SFB 1143 of the DFG.
S.\,R.~acknowledges support by the Australian Research Council (ARC) through Grant No. DP240100168.

 \


 \

\bibliography{refs_dec20}

\end{document}


\title{Supplemental Material --- Relevance of on-site and intersite Coulomb interactions
				     in the Kitaev-Heisenberg magnet Na$_3$Co$_2$SbO$_6$}

\author{Pritam Bhattacharyya}
\affiliation{Institute for Theoretical Solid State Physics, Leibniz IFW Dresden, Helmholtzstra{\ss}e~20, 01069 Dresden, Germany}
\affiliation{Department of Physics, Karpagam Academy of Higher Education, Coimbatore 641021, Tamil Nadu, India}
\affiliation{Centre for Computational Physics, Karpagam Academy of Higher Education, Coimbatore 641021, Tamil Nadu, India}

\author{Abdul Basit}
\affiliation{School of Physics, University of Melbourne, Parkville, VIC 3010, Australia}

\author{Thorben Petersen}
\affiliation{Institute for Theoretical Solid State Physics, Leibniz IFW Dresden, Helmholtzstra{\ss}e~20, 01069 Dresden, Germany}

\author{Stephan Rachel}
\affiliation{School of Physics, University of Melbourne, Parkville, VIC 3010, Australia}

\author{Satoshi Nishimoto}
\affiliation{Institute for Theoretical Solid State Physics, Leibniz IFW Dresden, Helmholtzstra{\ss}e~20, 01069 Dresden, Germany}
\affiliation{Department of Physics, Technical University Dresden, 01069 Dresden, Germany}

\author{Liviu Hozoi}
\affiliation{Institute for Theoretical Solid State Physics, Leibniz IFW Dresden, Helmholtzstra{\ss}e~20, 01069 Dresden, Germany}

\date\today
\maketitle

{\it Geometry and basis sets details.\,}
Atomic coordinates as reported by Viciu $et$ $al.$ \cite{Viciu_2007} were employed in our computations.
All quantum chemical computations were carried out using the {\sc molpro} suite of programs \cite{
Molpro}.
For each type of embedded cluster, the crystalline environment was modeled as a large array of point
charges which reproduces the crystalline Madelung field within the cluster volume;
we employed the {\sc ewald} program \cite{Klintenberg_et_al,Derenzo_et_al} to generate the point-charge
embeddings.

To clarify the Co-site multiplet structure, a cluster including one `central' CoO$_6$ unit was considered.
The quantum chemical study was initiated as CASSCF calculations \cite{olsen_bible,MCSCF_Molpro} with
active orbital spaces containing the five 3$d$ orbitals of the central Co ion.
Post-CASSCF correlation computations were performed at the MRCI level, with single and double excitations
\cite{olsen_bible,MRCI_Molpro} out of the Co 3$d$ and O 2$p$ orbitals of the central CoO$_6$ octahedron.
Spin-orbit couplings (SOCs) were accounted for following the procedure described in Ref.~\cite{SOC_Molpro}.
We employed all-electron triple-$\zeta$ basis sets (BSs) for the central Co site \cite{Co_basis} and
six adjacent O ligands \cite{O-ligands}.
The three transition-metal nearest neighbors were represented as closed-shell Zn$^{2+}$ ions, using
large-core pseudopotentials Zn ECP28MWB and uncontracted (3s2p) BSs \cite{Schautz1998};
the three in-plane adjacent Sb species were described by large-core pseudopotentials Sb ECP46MDF
and (4s4p)/[2s2p] BSs \cite{Sb_pseudo} from the {\sc molpro} library.
The outer 18 O ligands associated with the six adjacent octahedra were modeled through minimal
all-electron atomic-natural-orbital (ANO) BSs \cite{Pierloot1995}.
Large-core pseudopotentials were adopted for the 24 adjacent Na cations \cite{Fuentealba_Na}.

Clusters with two edge-sharing CoO$_6$ octahedra in the central region were considered in order to
derive the intersite effective magnetic couplings.
We utilized all-electron triple-$\zeta$ BSs for the `central' Co sites \cite{Co_basis}.
All-electron BSs of quintuple-$\zeta$ quality were employed for the two bridging ligands \cite{O-ligands}
while all-electron triple-$\zeta$ BSs were applied for the remaining eight O anions \cite{O-ligands}
associated with the two octahedra of the reference unit.
The four adjacent transition-metal ions were represented as closed-shell Zn$^{2+}$ species, using large-core
pseudopotentials Zn ECP28MWB and uncontracted (3s2p) BSs \cite{Schautz1998}, and the four adjacent Sb cations through
large-core pseudopotentials Sb ECP46MDF and (4s4p)/[2s2p] BSs \cite{Sb_pseudo}.
The outer 14 O ligands associated with the four adjacent SbO$_6$ octahedra were described through minimal
all-electron ANO BSs \cite{Pierloot1995}.
Large-core pseudopotentials were considered for the 24 Na nearby cations \cite{Fuentealba_Na}.

\begin{figure}[!b]
\centering
\includegraphics[width=0.8\linewidth]{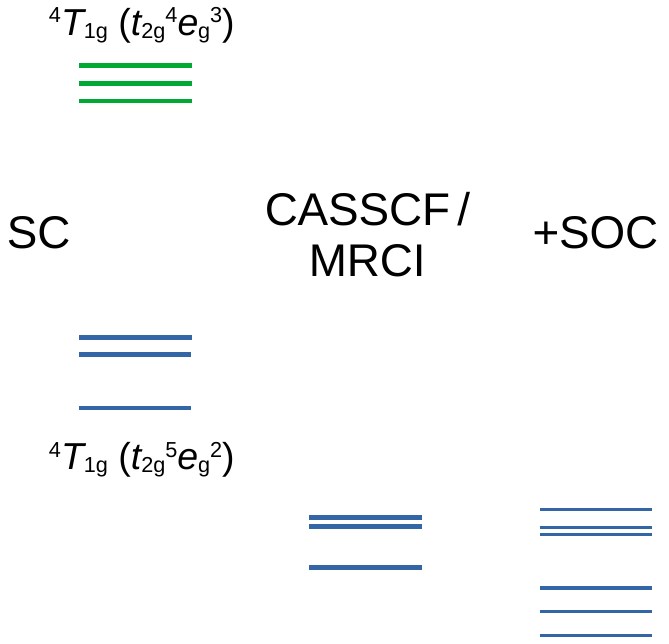}
\caption{
$^4T_{1g}$ terms providing the largest contributions to the ground-state wave-function.
CASSCF(+MRCI) accounts for $^4T_{1g}$($t_{2g}^5e_g^2$)\,--\,$^4T_{1g}$($t_{2g}^4e_g^3$) interaction
and additional, weaker configuration-interaction effects.
With spin-orbit interactions and noncubic environment, each $^4T_{1g}$ term evolves into a set of
six, distinct KDs.
Only the evolution of the lower $^4T_{1g}$ component is sketched, figure not to scale.
}
\label{fig_3d_levels}
\end{figure}

{\it On-site many-body physics.\,}
Effects of on-site configuration-interaction and SOC are sketched in Supplementary Figure
\ref{fig_3d_levels}:
there is substantial $^4T_{1g}$($t_{2g}^5e_g^2$)\,--\,$^4T_{1g}$($t_{2g}^4e_g^3$) interaction,
with an admixture of 8\% of the higher-lying configuration to the ground state, and additional
smaller contributions to the ground-state wave-function through both Coulomb and second-order
spin-orbit interactions (see also Table\;I in the main text).
Notations as in cubic symmetry are used, although low-symmetry crystal fields \cite{Viciu_2007}
split up the 3$d$ $t_{2g}$ (and $e_g$) sublevels.
In the presence of first-order SOC and noncubic fields, the $^4T_{1g}$ manifold entails a number
of six, distinct Kramers doublets (KDs).

\begin{figure}[!b]
\centering
\includegraphics[width=1\linewidth]{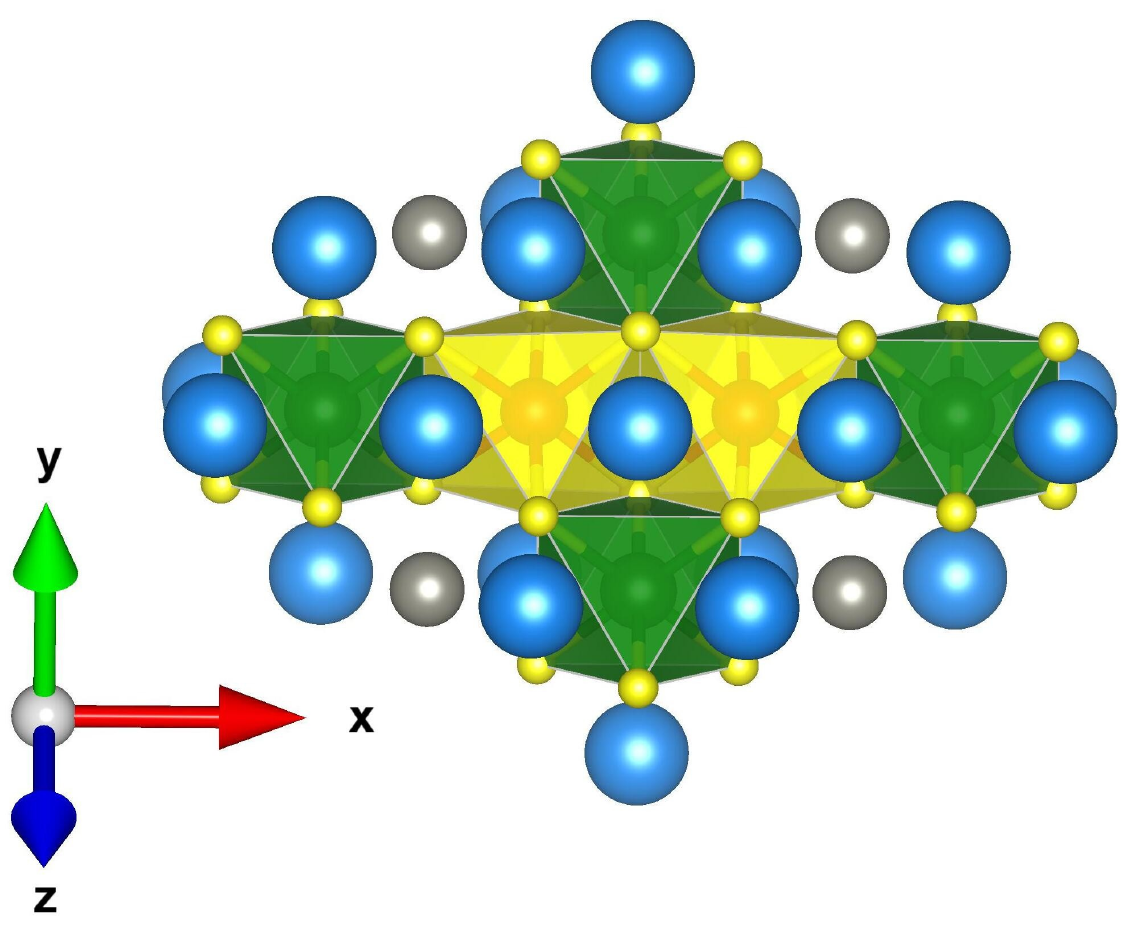}
\caption{
Cluster employed for deriving the nearest-neighbor effective magnetic couplings.
CoO$_6$ and SbO$_6$ octahedra are represented in yellow and green, respectively;
Na species are shown in blue.
The embedding is not pictured.
}
\label{fig_cluster}
\end{figure}

{\it Analysis of exchange contributions.\,}
CASSCF computations were carried out with ten valence orbitals (Co 3$d$) and 14 electrons
as active (abbreviated as (14e,10o) active space).
In the subsequent MRCI correlation treatment, single and double excitations out of the central-unit
Co 3$d$ and bridging O 2$p_z$ levels were considered.
We used the Pipek-Mezey methodology \cite{PM_method} to obtain localized central-unit orbitals.
Different from previous quantum chemical investigations (e.\,g., on RuCl$_3$ in ref.\;\cite{Yadav2016}),
where the core and semi-core orbitals were kept frozen at CASSCF level, as obtained from a preliminary
Hartree-Fock calculation preceding the CASSCF step, all orbitals were here reoptimized in the CASSCF
variational procedure.
Interestingly, for the particular case of RuCl$_3$, by full orbital optimization in CASSCF the sign
of the Heisenberg $J$ is reversed:
from $J\!=\!1.2$ meV in ref.\;\cite{Yadav2016}, we arrive at $J\!=-0.4$ meV in the final MRCI spin-orbit
computation if all orbitals are reoptimized in CASSCF.
The other nearest-neighbor coupling parameters are less affected, with $K\!=-3.7$, $\Gamma\!=\!1.5$,
and $\Gamma'\!=\!0.4$.

The CASSCF optimization was performed for the nine singlet, nine triplet, nine quintet, and nine septet
states associated with the (14e,10o) setting.
Those were the states for which SOCs were further accounted for \cite{SOC_Molpro}, at either SC, SSCAS,
CASSCF, or MRCI level, which finally yields a number of 144 spin-orbit states in each case.
The lowest four spin-orbit eigenstates from the {\sc molpro} output (with eigenvalues lower by $\sim$30
meV or more compared to other states, as illustrated in Table\;I, main article) were mapped onto the eigenvectors
of the effective spin Hamiltonian (Eq. 1, main article), following the procedure described in refs.~\cite{
Ir214_niko_15,Yadav2016}\,:\,
those four expectation values and the matrix elements of the Zeeman Hamiltonian in the basis of the
four lowest-energy spin-orbit eigenvectors (at either SC, SSCAS, CASSCF, or MRCI level) are put in
direct correspondence with the respective eigenvalues and matrix elements of Eq. 1, main article.

The SSCAS label denotes a CI calculation where only on-site intra-3$d$ excitations
were included; the CASSCF orbitals were employed (also for the SC numerical experiment).

\begin{figure}[!t]
\centering
\includegraphics[width=1.0\linewidth]{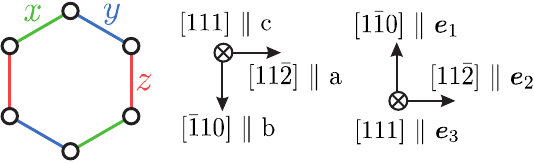}
\caption{
Hexagonal ring with three different types of Kitaev links (left), $a$,\,$b$,\,$c$ coordinates
(middle), and $(\hat{\boldsymbol{e}}_1,\hat{\boldsymbol{e}}_2,\hat{\boldsymbol{e}}_3)$
reference frame in $XYZ$ representation (right).
}
\label{fig:XXZ_frame}
\end{figure}

{\it Transformation to $XXZ$ frame.\,}
With uniaxial trigonal anisotropy, it may be convenient to rewrite the effective spin Hamiltonian in
$XXZ$ representation (see Supplementary Figure \ref{fig:XXZ_frame}), in terms of the local coordinates
\begin{align}
	\hat{\boldsymbol{e}}_1&=\frac{1}{\sqrt{2}}(\hat{\boldsymbol{x}}-\hat{\boldsymbol{y}}) \ ,\\
	\hat{\boldsymbol{e}}_2&=\frac{1}{\sqrt{6}}(\hat{\boldsymbol{x}}+\hat{\boldsymbol{y}}-2\hat{\boldsymbol{z}}) \ ,\\
	\hat{\boldsymbol{e}}_3&=\frac{1}{\sqrt{3}}(\hat{\boldsymbol{x}}+\hat{\boldsymbol{y}}+\hat{\boldsymbol{z}}) \ .
\end{align}
The components of the spin-$\frac{1}{2}$ operator can be expressed then as
\begin{align}
	\tilde{S}_i^1&=\frac{1}{\sqrt{2}}(S_i^x-S_i^y) \ ,
	\label{S1}\\
	\tilde{S}_i^2&=\frac{1}{\sqrt{6}}(S_i^x+S_i^y-2S_i^z) \ ,
	\label{S2}\\
	\tilde{S}_i^3&=\frac{1}{\sqrt{3}}(S_i^x+S_i^y+S_i^z) \ .
	\label{S3}
\end{align}
From Eqs.~\eqref{S1}-\eqref{S3}, the components of the original spin operator are
\begin{align}
	\tilde{S}_i^x&=\frac{1}{\sqrt{6}}(\sqrt{3}S_i^1+S_i^2+\sqrt{2}S_i^3) \ ,
	\label{Sx}\\
	\tilde{S}_i^y&=\frac{1}{\sqrt{6}}(-\sqrt{3}S_i^1+S_i^2+\sqrt{2}S_i^3) \ ,
	\label{Sy}\\
        \tilde{S}_i^z&=\frac{1}{\sqrt{3}}(-\sqrt{2}S_i^2+S_i^3) \ .
	\label{Sz}
\end{align}
Substituting Eqs.~\eqref{Sx}-\eqref{Sz} into the original Hamiltonian, for $z$-type magnetic bonds,
the corresponding $XXZ$ Hamiltonian becomes 
\begin{align}
	\label{eqn:zbondXXZ}
	\nonumber
	\mathcal{H}_{ij}^z = &J_{xy} (\tilde{S}_i^1\tilde{S}_j^1 + \tilde{S}_i^2\tilde{S}_j^2)
	+ 2J_{\pm\pm} (\tilde{S}_i^1\tilde{S}_j^1 - \tilde{S}_i^2\tilde{S}_j^2)\\
	&+ J_z \tilde{S}_i^3\tilde{S}_j^3 + J_{z\pm}(\tilde{S}_i^2\tilde{S}_j^3 + \tilde{S}_i^3\tilde{S}_j^2) \ ,
\end{align}
where $J_{xy}=J+K/3-\Gamma/3-2\Gamma'/3$, $J_z=J+K/3+2\Gamma/3+4\Gamma'/3$,
$J_{\pm\pm}=-K/3-2\Gamma/3+2\Gamma'/3$, and 
$J_{z\pm}=-\sqrt{2}K/3+\sqrt{2}\Gamma/3-\sqrt{2}\Gamma'/3$.

Similarly, for $x$ and $y$ bonds, starting from the initial expressions
\begin{align}
 	\label{eqn:zbondorg}
 	\nonumber
 	\mathcal{H}_{ij}^x = &J{\bf{\tilde{S}}}_i\cdot {\bf{\tilde{S}}}_j + 
 	K\tilde{S}_i^x\tilde{S}_j^x + \Gamma (\tilde{S}_i^y\tilde{S}_j^z + \tilde{S}_i^z\tilde{S}_j^y)\\
 	&+ \Gamma'(\tilde{S}_i^z\tilde{S}_j^x + \tilde{S}_i^x\tilde{S}_j^z + \tilde{S}_i^x\tilde{S}_j^y
 	+ \tilde{S}_i^y\tilde{S}_j^x) \ ,\\
 	\label{eqn:zbondorg} 
\nonumber
\mathcal{H}_{ij}^y = &J{\bf{\tilde{S}}}_i\cdot {\bf{\tilde{S}}}_j + 
K\tilde{S}_i^y\tilde{S}_j^y + \Gamma (\tilde{S}_i^z\tilde{S}_j^x + \tilde{S}_i^x\tilde{S}_j^z)\\
&+ \Gamma'(\tilde{S}_i^x\tilde{S}_j^y + \tilde{S}_i^y\tilde{S}_j^x + \tilde{S}_i^y\tilde{S}_j^z
+ \tilde{S}_i^z\tilde{S}_j^y) \ ,
\end{align}
by substituting Eqs. \eqref{Sx}-\eqref{Sz}, we obtain
\begin{align}
 	\label{eqn:xbondXXZ}
 	\nonumber
 	\mathcal{H}_{ij}^y &= (J+\frac{K}{2}-\Gamma')\tilde{S}_i^1\tilde{S}_j^1
  	+(J+\frac{K}{6}-\frac{2\Gamma}{3}-\frac{\Gamma'}{3})\tilde{S}_i^2\tilde{S}_j^2\\
  	\nonumber	&+(J+\frac{K}{3}+\frac{2\Gamma}{3}+\frac{4\Gamma'}{3})\tilde{S}_i^3\tilde{S}_j^3\\
  	\nonumber
 	&-\frac{\sqrt{3}}{6}(K+2\Gamma-2\Gamma')(\tilde{S}_i^1\tilde{S}_j^2+\tilde{S}_i^2\tilde{S}_j^1)\\
 	\nonumber
 	&+\frac{\sqrt{2}}{6}(K-\Gamma+\Gamma')(\tilde{S}_i^2\tilde{S}_j^3+\tilde{S}_i^3\tilde{S}_j^2)\\
 	&-\frac{\sqrt{6}}{6}(K-\Gamma+\Gamma')(\tilde{S}_i^3\tilde{S}_j^1+\tilde{S}_i^1\tilde{S}_j^3) \ ,\\
 	\label{eqn:ybondXXZ}
\nonumber
\mathcal{H}_{ij}^x &= (J+\frac{K}{2}-\Gamma')\tilde{S}_i^1\tilde{S}_j^1
+(J+\frac{K}{6}-\frac{2\Gamma}{3}-\frac{\Gamma'}{3})\tilde{S}_i^2\tilde{S}_j^2\\
\nonumber	&+(J+\frac{K}{3}+\frac{2\Gamma}{3}+\frac{4\Gamma'}{3})\tilde{S}_i^3\tilde{S}_j^3\\
\nonumber
&+\frac{\sqrt{3}}{6}(K+2\Gamma-2\Gamma')(\tilde{S}_i^1\tilde{S}_j^2+\tilde{S}_i^2\tilde{S}_j^1)\\
\nonumber
&+\frac{\sqrt{2}}{6}(K-\Gamma+\Gamma')(\tilde{S}_i^2\tilde{S}_j^3+\tilde{S}_i^3\tilde{S}_j^2)\\
&+\frac{\sqrt{6}}{6}(K-\Gamma+\Gamma')(\tilde{S}_i^3\tilde{S}_j^1+\tilde{S}_i^1\tilde{S}_j^3) \ .
\end{align}



\begin{figure}[!hb!]
\centering
\includegraphics[width=0.8\linewidth]{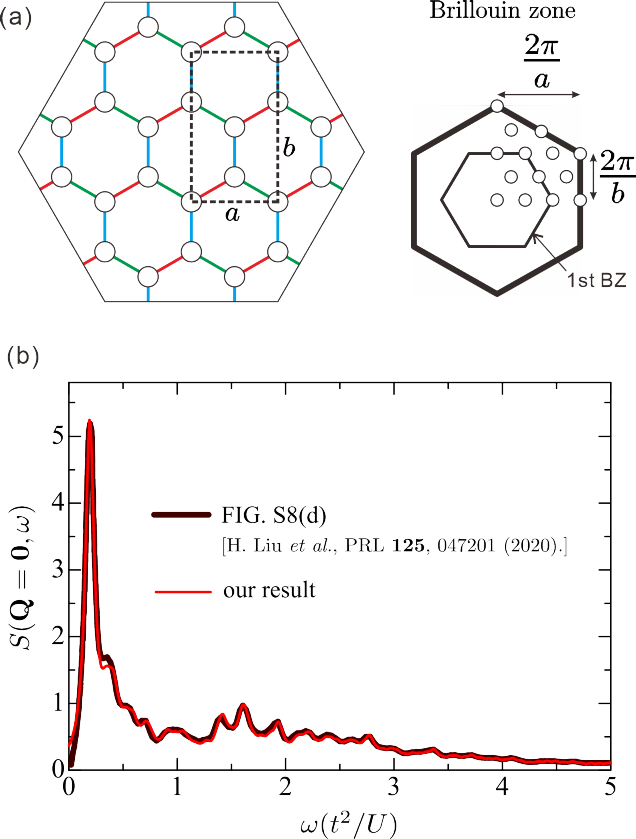}
\caption{
(a) 24-site cluster used for Lanczos ED with periodic boundary conditions
and the corresponding momenta in the extended Brillouin zone.
(b) Comparison of our dynamical structure factor at $\vec{Q}=\vec{0}$ with
the result reported in Ref.~\cite{Liu_2020}.
}
\label{fig:benchmark}
\end{figure}

{\it Exact-diagonalization analysis of the dynamical structure factor.\,}
To examine the validity of linear spin-wave theory (LSWT) for the present frustrated Kitaev system,
we employed Lanczos exact-diagonalization (ED) computations for a cluster with $N\!=\!24$ sites and
periodic boundary conditions.
The dynamical spin structure factor is defined as
\begin{eqnarray}
	\nonumber
	S^{\gamma\bar{\gamma}}(\vec{Q},\omega) &=& \frac{1}{\pi}\,{\rm Im} 
	\langle \psi_0 | (S^\gamma_{\vec{Q}})^\dagger 
	\frac{1}{\hat{H}+\omega-E_0-{\rm i}\eta} 
	S^\gamma_{\vec{Q}} | \psi_0 \rangle \\
	&=& \sum_\nu 
	|\langle \psi_\nu | S^\gamma_{\vec{Q}} | \psi_0 \rangle|^2 
	\delta(\omega - E_\nu + E_0),
	\label{spec}
\end{eqnarray}
where $\gamma = z,\,\pm$, and
$|\psi_\nu\rangle$ and $E_\nu$ denote the $\nu$th eigenstate and eigenenergy of the system
($\nu=0$ corresponds to the ground state).
The momentum-resolved spin operators are defined as
\begin{align}
	S^\gamma_{\vec{Q}} = \frac{1}{\sqrt{N}} 
	\sum_{l} e^{i\vec{Q}\cdot\vec{R}_l} S^\gamma_{\vec{R}_l},
	\label{operator_pbc}
\end{align}
where $\vec{R}_l$ runs over all lattice sites (see Supplementary Fig.~\ref{fig:benchmark}(a)).

For validation of our Lanczos implementation, we evaluated
\(
S(\vec{Q}=\vec{0},\omega)
= \frac{1}{2}\left[ S^{+-}(\vec{Q},\omega)
+ S^{-+}(\vec{Q},\omega) \right]
+ S^{zz}(\vec{Q},\omega),
\)
and confirm that the resulting low-energy spectrum agrees with that reported in Ref.~\cite{Liu_2020}
(see Supplementary Fig.~\ref{fig:benchmark}(b)).
In this benchmark calculation, we used the same model parameters as in Ref.~\cite{Liu_2020}, and the
broadening factor was set to $\eta\!=\!0.05\,t^2/U$.
These benchmark checks ensure that the ED results presented in this Supplemental Material are fully
consistent with previous studies.

Although the number of accessible momentum points is limited due to the finite cluster size, the ED
results agree reasonably well with the LSWT dispersions, as shown in Supplementary Fig.~\ref{fig:ED_LSWT}.
In particular, the dominant low-energy peaks follow the same trend as the magnon branches predicted
by LSWT.
This qualitative consistency indicates that LSWT captures the leading short-wavelength magnonlike
features of the ordered phase, despite strong quantum fluctuations expected beyond the harmonic
approximation.
The ED spectra consist of discrete levels and do not exhibit the broad continua characteristic of
fractionalized excitations in Kitaev magnets.
Therefore, the ED results should be regarded only as a \textit{consistency check}, rather than a
quantitative test of LSWT.
The qualitative agreement nevertheless supports the use of LSWT as baseline description for the magnon
contributions discussed in the main text.

\begin{figure}[t]
\centering
\includegraphics[width=0.8\linewidth]{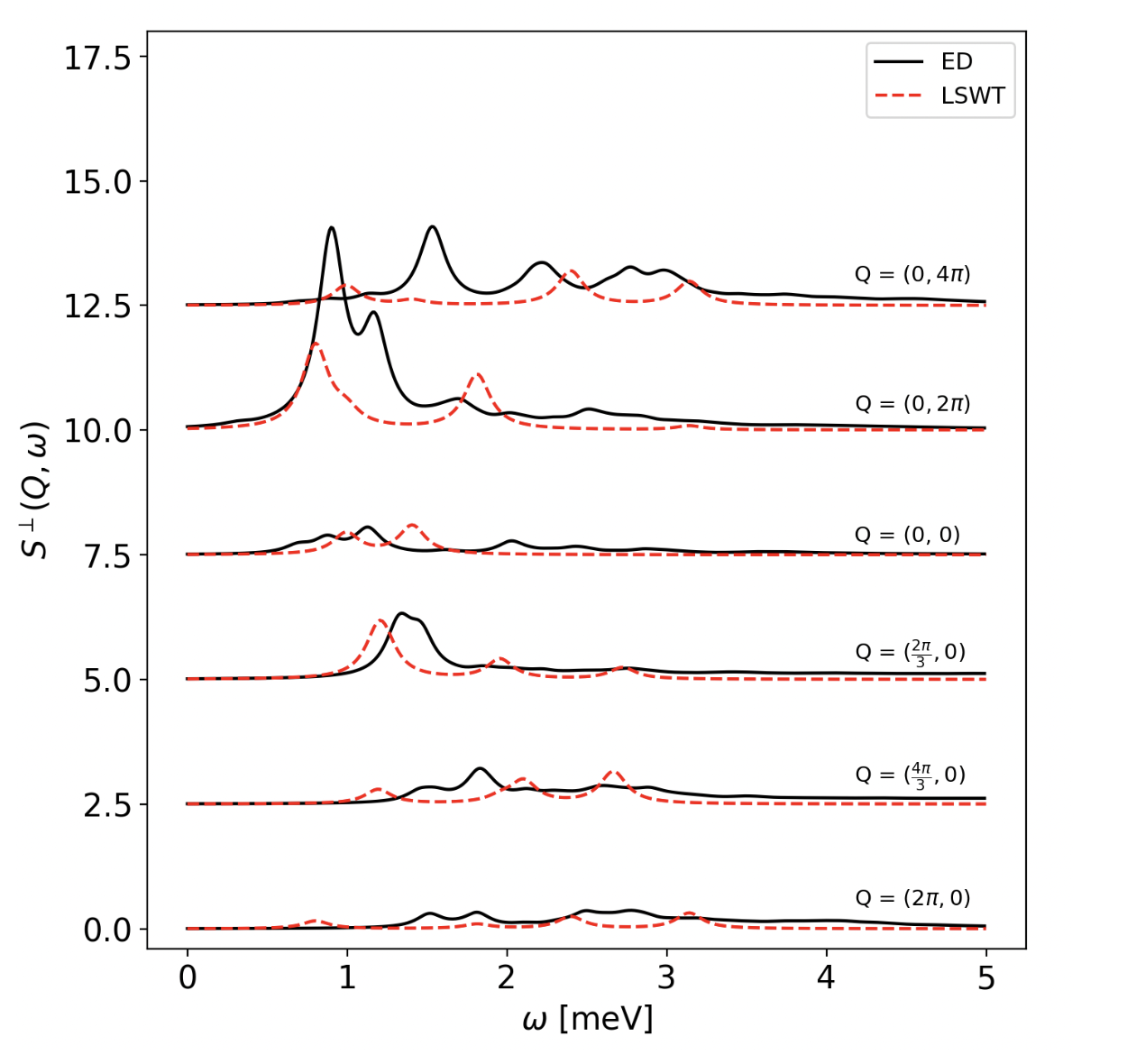}
\caption{
Comparison of the dynamical structure factors $S(\vec{Q},\omega)$ at
various momenta obtained from LSWT and Lanczos ED.
The spectra are convoluted with a Lorentzian broadening $\eta = 0.1\,\mathrm{meV}$.
}
\label{fig:ED_LSWT}
\end{figure}

\bibliography{refs_dec20}